\documentclass[a4paper,twocolumn,aps,pra,nosuperscriptaddress]{revtex4-1}

\usepackage[T1]{fontenc}
\usepackage[latin9]{inputenc}
\usepackage{color}
\usepackage{amsmath}
\usepackage{amssymb}
\usepackage{graphicx}
\usepackage[unicode=true,bookmarks=false,breaklinks=false,pdfborder={0 0 0},pdfborderstyle={},backref=false,colorlinks=true,pdfpagemode=FullScreen]
{hyperref}
\hypersetup{colorlinks,citecolor=blue,filecolor=black,linkcolor=red,urlcolor=blue}
\usepackage{breakurl}



\begin{document}

\title{Proposal for Anderson localization in transverse spatial degrees
of freedom of photons}

\author{Rafael M. Gomes}

\affiliation{Instituto de Física, Universidade Federal de Goiás, 74.690-900, Goiânia,
Goiás, Brazil}

\author{Wesley B. Cardoso}

\affiliation{Instituto de Física, Universidade Federal de Goiás, 74.690-900, Goiânia,
Goiás, Brazil}

\author{Ardiley T. Avelar}

\affiliation{Instituto de Física, Universidade Federal de Goiás, 74.690-900, Goiânia,
Goiás, Brazil}
\begin{abstract}
We propose an experimental setup for studying the Anderson localization
of light in the continuous transverse spatial degrees of freedom of
the photons. This physical phenomenon can be observed in the transverse
profile of a paraxial and quasi-monochromatic beam of light using
a spatial light modulator. The light modulator acts in the laser beam
as a weak random potential. Here, differently from the standard models
studied in the literature, our setup splits the dispersion and potential
terms along the beam evolution. By numerical simulations we confirm
the feasibility of our experimental proposal.
\end{abstract}
\maketitle

\section{Introduction}

Anderson localization - the phenomenon of transport suppression of
wave due to a destructive interference of the many paths associated
with coherent multiple scattering from the modulations of a disordered
potential - continues fascinating researchers since the appearance
of Anderson\textquoteright s 1958 paper \citep{Anderson58}. This
effect is a characteristic of wave physics and occurs when the disordered
potential presents a weak amplitude, making the localized state with
exponentially decaying tails and absence of diffusion. The Anderson
localization has been experimentally demonstrated in many scenarios
such as Bose-Einstein condensate (BEC) of atoms \citep{Wiersma97,Storzer06},
in 2D \citep{Billy08} and 1D \citep{Roati08} disordered photonic
lattices. From the theoretical viewpoint, the Anderson localization
has been studied in synthetic photonic lattice with random coupling
\citep{Pankov19}, in biological nanostructures of native silk \citep{Choi18},
on the surface plasmon polariton \citep{Petracek18}, and in Bose--Einstein
condensate with a weakly positive nonlinearity under the influence
of chaotic potentials \citep{Cardoso12NARWA,Cardoso2016b}. 

Potential applications ranging from solar cells to endoscopic fibers
have also motivated the investigation of the Anderson localization
of light that arisen by the understanding that the localization is
ubiquitous to all wave systems, in particular it should be present
in optics, where there is the remarkable analogy between the paraxial
equation for electromagnetic waves and the Schrödinger equation ruling
the quantum phenomena \citep{Soukoulis01,Wiersma13}. It is worth
to mention that light in optical domain furnishes an ideal system
to investigate localization effects since we have preserved coherence
and non-interacting bosons (photons) in order to satisfy the two required
assumptions of Anderson's model: time-invariant potential and absence
of interaction \citep{Segev13}.

In this context, transversal Anderson localization of light - firstly
proposed in \citep{Abdullaev80} and rediscovered a decade later in
\citep{Raedt89} - has started a promisor experimental scenario for
studying localization effects. It occurred due mainly to the experimental
observation of discrete transversal solitons \citep{Fleischer03}
that has opened the way to investigate localization effects in paraxial
disordered photonic systems, culminating with the remarkable experimental
observations of the transverse Anderson localization of light in 3D
random media \citep{Schwartz07,Lahini08}. In addition, the experimental
achievement of \citep{Segev13} in a two-dimensional photonic lattice
with random refractive index fluctuations induced on a photorefractive
crystal using an optical interference pattern, together large refractive
index available in optical fiber, becomes possible to glimpse transverse
Anderson localization as an effective waveguiding mechanism for the
light in random optical fiber \citep{Karbasi12}.

On the other hand, the transverse spatial continuous variables associated
with paraxial beam profile provide a robust and fertile testing scenario
for investigations of fundamentals of quantum mechanics related to
the EPR paradox \citep{Walborn11PRL}, quantum entanglement \citep{Tasca08,Lorenzo09},
entanglement beyond Gaussian states of light \citep{Gomes09PANS,Tasca13},
production of purely nonlocal optical vortex \citep{Gomes09PRL,Gomes11},
quantum information \citep{Walborn06PRL,Walborn08PRA,Almeida05},
quantum computation \citep{Tasca11}, and simulation of chaotic \citep{Lemos12}
and relativistic system such as Dirac equation \citep{Silva19}. Here,
taking advantageus of the facilities available in linear paraxial
optics scenario, we propose the experimental realization of the quantum
Anderson Localization dynamics in the spatial degrees of freedom of
the photons in a monochromatic paraxial light. To this end, a spatial
light modulator (SLM) is used to implement a potential in the transverse
profile of the field propagating freely, resulting to Anderson Localization
in the transverse position of the photon.

The paper is structured as follows. In Sec. \ref{TCV} we discuss
the transversal continuous variables. The experimental setup is proposed
and detailed in Sec. \ref{ES}. In Sec. \ref{NR}, we present the
results of a numerical simulation, showing the feasibility of the
proposal. Finally, we present our conclusions and final remarks in
Sec. \ref{CFR}.

\section{Transverse Continuous Variables \label{TCV}}

We propose an optical physical system to implement Anderson localization
in the transverse profile of a laser beam. The complex amplitude of
the electric component of a electromagnetic field can be represented
by $\vec{E}(x,y,z)=A(x,y,z)\vec{\epsilon}e^{-ikz}$, where $\vec{\epsilon}$
is the polarization vector, $k=2\pi/\lambda$ is the wavenumber and
$A(x,y,z)$ is the complex envelope. Consider $A(x,y,z)$ a function
that varies slowly on the neighborhood of wavelength $\lambda$, such
that the complex envelope of a monochromatic laser beam satisfies
the paraxial Helmholtz equation \citep{Saleh07}, given by
\begin{equation}
\frac{i}{k}\frac{\partial}{\partial z}A(x,y,z)=-\frac{1}{2k^{2}}\left(\frac{\partial^{2}}{\partial x^{2}}+\frac{\partial^{2}}{\partial y^{2}}\right)A(x,y,z).\label{eq.1}
\end{equation}
The above equation behaves like a bidimensional, \emph{alias} (2+1),
Schrödinger equation for the complex monochromatic field $A(x,y,z)$
in free space, with the propagation variable $z$ of the field being
analogous to the time in the standard Schrödinger equation while $x$
and $y$ are the transverse positions of the field. It is known that
Eq. (\ref{eq.1}) admits a Gaussian solution representing the transversal
profile of the laser beam. Without loss of generality, one can study
the system in only one spatial transverse dimension, and from Eq.
(\ref{eq.1}) obtains the form
\begin{equation}
\frac{i}{k}\frac{\partial}{\partial z}\phi(x,z)=-\frac{1}{2k^{2}}\frac{\partial^{2}}{\partial x^{2}}\phi(x,z),\label{eq.2}
\end{equation}
where $\phi(x,z)$ is the component of the transversal profile in
$x$ direction. Note that, if one considers the variable $z$ as a
time variable and replace the wavenumber $k$ to $1/\hbar$, the Eq.
(\ref{eq.2}) becomes completely analogous to the standard one-dimensional
Schrödinger equation for the field $\phi(x,z)$ in free space for
a particle of unitary mass. By this analogy, David Stoler introduced
the isomorphism between the Hilbert space in the transverse variables
of single photons with the Hilbert space of the nonrelativistic quantum
state of single point particle. The operators $\boldsymbol{x}$ and
$\boldsymbol{\theta}_{x}$ are defined as canonical conjugate operators
that obey the commutation relations $[\boldsymbol{x},\boldsymbol{\theta}_{x}]=i/k=i\hbar$
or $[\boldsymbol{x},\boldsymbol{q}_{x}]=1$, where the dimensional
transverse momentum is $q_{x}=\frac{1}{i}\partial/\partial x=\frac{1}{i}\theta_{x}k$
\citep{Stoler81,Nienhuis93}.

\section{Experimental setup \label{ES}}

\begin{figure}[tb]
\centering \includegraphics[width=1\columnwidth]{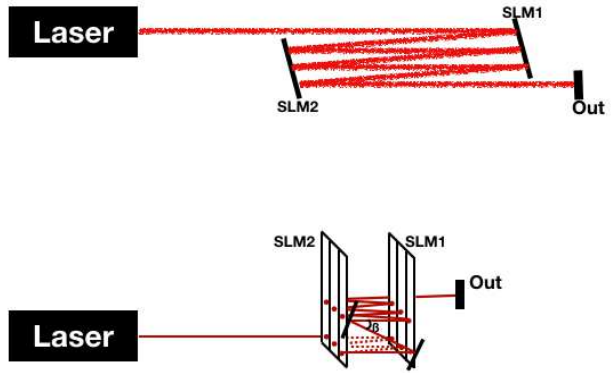}

\caption{Experimental setup to implement the Anderson Localization of the transverse
profile of the laser beam. The upper figure represents the upper perspective
of the setup. The bottom figure represents the lateral view of the
setup configuration. The two SLM's (SLM1 and SLM2) is configured with
the same phase. Note in the bottom figure that the number of the times
that the beam reflect in the SLM's is related to the input $\beta$
angle.}

\label{fig.1}
\end{figure}

In our experimental proposal, we consider the use of transverse profile
of a quasi-monochromatic and paraxial light beam and two SLMs properly
positioned so that the beam suffers multiple reflections, each one
of them with the same random phase impression. This experimental setup
shown in Fig. \ref{fig.1} allows us to observe the Anderson localization
in the transversal profile of the photons.

In the free propagation path the amplitude of the light beam is well
described by the Eq. (\ref{eq.2}). Indeed, in Fresnel approximation
the evolution operator can be represented by the transfer function
\citep{Saleh07}
\begin{equation}
U(q_{x})=e^{-izq_{x}^{2}/2k}.\label{eq:3}
\end{equation}
Following, the interaction of the light beam with the SLM leads to
the field suffers a phase changing according to the transformation
$e^{if(x)}$. In order to observe the Anderson localization of the
photons, one need to use a random profile function $f(x)$. Then,
the transformation in the complex envelope of the field after a reflection
in the SLM and the propagation in free space by a distance $d$, before
the next reflection, it is represented by the evolution operator
\begin{equation}
U'=e^{-idq_{x}^{2}/2k}e^{if(x)}.\label{eq:4}
\end{equation}
Consequently, after $N$ reflections in the SLM's, as represented
by the setup in Fig. \ref{fig.1}, the resulting transfer function
is given by
\begin{equation}
U''=\prod_{j=1}^{N}e^{-idq_{x}^{2}/2k}e^{if(x)}.\label{eq:5}
\end{equation}

On the other hand, the wavefunction $\phi$ of a massive point particle
subject to the potential $V(x)$ is well described by the Schrödinger
equation
\begin{equation}
i\hbar\frac{\partial\phi}{\partial t}=-\frac{\hbar^{2}}{2m}\frac{\partial^{2}\phi}{\partial x^{2}}+V(x)\phi,\label{1DSE}
\end{equation}
whose evolution operator is given by $U=\exp\{-i[p^{2}/2m+V(x)]t/\hbar\}$.
In numerical methods, the split-step method it is commonly employed
to solve the Eq. (\ref{1DSE}). The main idea of this method is to
split the evolution operator $U$ into the time interval $\Delta t$
in the form
\begin{equation}
U=\exp\{-i[p^{2}/2m]\Delta t/\hbar\}\exp\{-i[V(x)]\Delta t/\hbar\},\label{OESE}
\end{equation}
by using the well known Baker-Campbell-Hausdorff formula. The precision
of this approximation depends on the value of $\Delta t$. In fact,
the smaller the time interval $\Delta t$ is greater the accuracy
of the approximation is. We emphasize that the result obtained in
Eq. (\ref{OESE}) is similar to that of Eq. (\ref{eq:4}). Then, since
the Eq. (\ref{1DSE}) admits the Anderson localization to appropriate
random potentials $V(x)$, we expect a similar behavior to the transverse
profile of the light beam in the experimental setup depicted in Fig.
\ref{fig.1}.

\section{Numerical results \label{NR}}

\begin{figure}[tb]
\includegraphics[width=1\columnwidth]{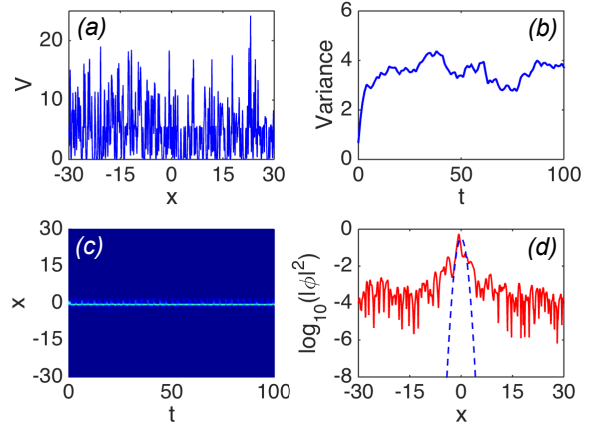}

\caption{(a) Phase printing function $V(x)$ modeled by a set of 300 identical
spikes randomly distributed along the $x$ axis according to the Eqs.
(\ref{eq2}) and (\ref{eq3}). (b) The variance of the the localized
state $\sqrt{\langle x^{2}\rangle-\langle x\rangle^{2}}$ versus the
evolution time, \emph{alias} the longitudinal propagation coordinate
$z$ of the light beam. (c) The light beam profile ($|\phi|^{2}$)
by considering a Gaussian input state in the real time evolution.
(d) Light beam profile of a typical configuration in log scale obtained
numerically from the Eq. (\ref{eq:5}) at $t=100$ (or equivalently
$10^{4}$ interactions of the beam with the SLM) in solid (red) curve
and the input Gaussian profile in dashed curve.}

\label{fig.2}
\end{figure}

In the experimental setup shown in Fig. \ref{fig.1}, the spatial
light modulators SLM1 and SLM2 are responsible to create a random
profile for the phase print $V(x)\equiv f(x)/dt$, acting as a potential
in Schrödinger-like equation (\ref{eq.2}), with $dt$ being a small
parameter. As an example, in Fig. \ref{fig.2}(a) we display a typical
potential profile obtained via a superposition of 300 speckles generated
in random positions. A detailed description, including a algorithm
for an experimental generation of speckle patterns of light can be
found in Ref. \citep{Huntley89}.

To implement the speckle patterns similar to that of Ref. \citep{Huntley89}
in the SLM, we construct an algorithm based on the Ref. \citep{Cheng10}.
To this end, the phase printing function $V(x)$ can be modeled by
a set of $S$ identical spikes randomly distributed along the $x$
axis \citep{Sanchez-Palencia08}
\begin{eqnarray}
V(x)=V_{0}\sum_{j=1}^{S}v(x-x_{j}),\label{eq2}
\end{eqnarray}
with $V_{0}$ being the strength of the spike, and $v(x-x_{j})$ are
the profile associated to a single spike at random position $x_{j}$.
We assume Gaussian spikes given by \citep{Sanchez-Palencia08} 
\begin{eqnarray}
v(x)=\left(\sigma\sqrt{\pi}\right)^{-1}\exp\left(-\frac{x^{2}}{\sigma^{2}}\right),\label{eq3}
\end{eqnarray}
with a specific width $\sigma$. The statistical average of the disordered
potential (\ref{eq2}) and the auto-correlation function for the phase
printing function $V(x)$ are given by $\langle V(x)\rangle\equiv\int_{-L}^{L}V(x)dx/2L=V_{0}/D$
and $C(d)=\langle V(x)V(x+d)\rangle-\langle V(x)\rangle^{2}$, respectively,
with $D$ being the average spacing between spikes in the spatial
extension $2L$. Also, the average spike height $V_{S}$ is defined
by $V_{S}=\left[\int_{-L}^{L}dx\left(V(x)-\langle V(x)\rangle\right)^{2}/2L\right]^{1/2}$.
As a typical example we set $S=300$, $L=30$, $V_{0}=1$ and a small
width $\sigma=0.1$. 

The numerical simulations were performed by splitting the dispersion
term due to the free propagation of the field and the phase printing
generated as effect of the spatial light modulators under this field.
Indeed, this is truly physical mechanism involved in the present protocol,
once these effects does not occurs simultaneously.

In Fig. \ref{fig.2}(b) we display the variance of the the localized
state $\sqrt{\langle x^{2}\rangle-\langle x\rangle^{2}}$ versus the
evolution time, \emph{alias} the longitudinal propagation coordinate
$z$ of the light beam. The parameters used in our simulations were
in a such way that the time interval $\Delta t=1$ corresponds to
100 reflections of the light beam by the SLMs. As a result, the absence
of variance growth implies the state localization. In order to evidence
the result shown in Fig. \ref{fig.2}(b), we display in Fig. \ref{fig.2}(c)
the light beam profile ($|\phi|^{2}$) by considering a Gaussian input
state in the real time evolution up to a generic time $t=100$, corresponding
to $10^{4}$ interactions of the beam with the SLMs.

Finally, in Fig. \ref{fig.2}(d) we consider the light beam profile
of a typical configuration in log scale obtained numerically from
the Eq. (\ref{eq:5}) at $t=100$. One can observe from this plot
that the tails of the output state (solid line) decay much more smoothly
(almost linear on average) than that for the Gaussian state, shown
by the dashed curve in Fig. \ref{fig.2}(d). This is the main evidence
of the Anderson localization \citep{Wiersma97,Storzer06,Schwartz07,Lahini08,Billy08,Roati08}.

\section{Conclusion and Final Remarks \label{CFR}}

We propose here a new experimental platform to study the Anderson
localization in the transverse degrees of freedom of photons, employing
SLM\textquoteright s to print a random phase in the transversal profile
of a laser beam. We confirmed the presence of Anderson localization
in the transverse degree of freedom of photons by direct numerical
simulations. The simplicity of the experimental configuration proposed
here to control the disordered potential and to detect the transversal
intensity of the beam is a prime advantage of the work. The tuning
of the disordered potential continues to be one of the prime challenges
of the experimental observation of the Anderson Localization in others
scenarios. On the other hand, the spatial light modulator absorbs
2\% of the beam intensity at each reflection, attenuating the beam
according to the number of reflections. However, this can be contoured
using avalanche photodetectors \citep{Gomes09PANS}. Furthermore,
the increasing technological development of more efficient SLM\textquoteright s
will enable to observe the resulting beam of light in CCD cameras.
Finally, we hope that this work opens new possibilities to study the
Anderson localization phenomenon in quantum optics.

\subsection*{Acknowledgments}

We acknowledge acknowledge financial support from the Brazilian agencies
CNPq (\#304073/2016-4,  \#425718/2018-2 \& \#479956/2013-8), CAPES,
and FAPEG (PRONEM \#201710267000540, PRONEX \#201710267000503). This
work was performed as part of the Brazilian National Institute of
Science and Technology for Quantum Information (INCT-IQ).

%



\end{document}